\documentclass[journal]{IEEEtran}

\usepackage[T1]{fontenc}

\usepackage{amsmath}
\usepackage{graphicx}
\usepackage{caption}
\usepackage{tabularx,multirow,booktabs,lipsum}
\usepackage{subfig}
\usepackage{color}
\usepackage{float} 
\usepackage{paralist} 
\usepackage[T1]{fontenc}
\usepackage{currvita}
\usepackage{epstopdf}
\usepackage{amsmath}
\usepackage{amsfonts}
\usepackage{amssymb}

\ifCLASSINFOpdf
 \else
\fi

\usepackage{amsmath}
\usepackage{mathrsfs}
\usepackage{amsfonts}
\usepackage{amssymb}

\usepackage[cmintegrals]{newtxmath}

\usepackage{algorithm}
\usepackage[noend]{algpseudocode}

\DeclareMathAlphabet\mathbfcal{OMS}{cmsy}{b}{n}

\makeatletter
\def\BState{\State\hskip-\ALG@thistlm}
\makeatother

\begin{document}

\title{Software-Defined Microgrid Control for Resilience Against Cyber Attacks}

\author{Pietro Danzi,~\IEEEmembership{Student Member,~IEEE,} Marko Angjelichinoski,~\IEEEmembership{Student Member,~IEEE,} \v{C}edomir Stefanovi\'c,~\IEEEmembership{Senior Member,~IEEE,} Tomislav Dragi\v{c}evi\'c,~\IEEEmembership{Member,~IEEE,} Petar Popovski,~\IEEEmembership{Fellow,~IEEE}
\thanks{P. Danzi, M. Angjelichinoski, \v{C}. Stefanovi\'c and P. Popovski are with the Department of Electronic Systems, Aalborg University, Denmark, Email: \{pid,maa,cs,petarp\}@es.aau.dk. Tomislav Dragi\v{c}evi\'c is with the Department of Energy Technology, Aalborg University, Denmark, Email: tdr@et.aau.dk.}
\thanks{The work presented in this paper was supported in part by EU, under grant agreement no. 607774 ``ADVANTAGE".}}

\markboth{submitted to IEEE Transactions on Smart Grid (2018)}%
{Danzi \MakeLowercase{\textit{et al.}}: Software-Defined Microgrid Control for Resilience Against Cyber Attacks}
\maketitle

\begin{abstract}
Microgrids (MGs) rely on networked control supported by off-the-shelf wireless communications. This makes them vulnerable to cyber-attacks, such as denial-of-service (DoS). 
In this paper, we mitigate those attacks by applying the concepts of (i)~separation of data plane from network control plane, inspired by the software defined networking (SDN) paradigm, and (ii)~agile reconfiguration of the data plane connections.
In our architecture, all generators operate as either voltage regulators (active agents), or current sources (passive agents), with their operating mode being locally determined, according the global information on the MG state.
The software-defined MG control utilizes the fact that, besides the data exchange on the wireless channel, the power-grid bus can be used to create side communication channels that carry control plane information about the state of the MG.
For this purpose, we adopt power talk, a modem-less, low-rate, power-line communication designed for direct current (DC) MGs.
The results show that the proposed software-defined MG offers superior performance compared to the static MG, as well as resilience against cyber attacks.
\end{abstract}


%
\IEEEpeerreviewmaketitle

\section{Introduction}

\IEEEPARstart{M}{icrogrids} (MGs) are key components in the emerging distributed Smart Grid landscape. They are defined as localized and self-sustainable clusters of Distributed Energy Resources (DERs) and loads, interconnected via power distribution network through power electronic converters (PECs) 
~\cite{meng2017review}.
Direct-current (DC) MGs are gaining popularity due to absence of reactive power, synchronization issues, as well as natural fit with emerging DC loads, energy storage systems and renewable sources~\cite{ref1m}.

The widespread adoption of the MG paradigm is attributed to the development of robust, secure, and non-supervised control systems, which are software-implemented and executed by PECs~\cite{ref2m}.
Hence, PECs take on the role of \emph{control agents} that perform the local sampling of the grid state and the overall MG regulation.
The control performed by PECs is usually hierarchically organized to distinguish among different objectives~\cite{ref01}. 
Specifically, the MG control architecture consists of three levels. 
The primary control is localized within the agent and guarantees the MG stability, while the secondary/tertiary control require cooperation among the agents in order to restore the voltage and optimize the power flow, respectively.

The cooperation among agents is supported by an external communication network, which can experience delay and packet loss, thereby deteriorating the MG operation. With non-ideal links, the secondary/tertiary control operate in a suboptimal regime. 
A possible solution is the adoption of robust control schemes that are tolerant against communication delays and/or packet drops, cf.~\cite{ref02}.
However, when the communication network is attacked by e.g. Denial of Service (DoS), the secondary/tertiary control fails~\cite{ref03}.
In addition to the poor adaptation to communication impairments, another deficiency of existing MG control is the information security: the increase of the number of attacks against big-scale power systems raises a concern also about the cyber-resilience of the small scale grids.
In fact, even if the MG failure does not lead to large blackouts, it may damage sensitive loads, such as electronic equipment, or cause the tripping of DERs~\cite{che2014adaptive}.

The concerns described above, as well as the possibility to abstract the agents' functionalities from the PECs' hardware, establish the motivation to redesign the multi-agent MG control towards a more reconfigurable and secure system.
Unlike the existing works, the role of each agent in our proposal is not statically defined: thanks to software reconfiguration, a PEC can operate as an active control agent as well as passively act as a sensor that delegates the control to active agents.
The operation mode is selected according to the MG status and to the status of the external communication network.
The resulting architecture, first introduced in~\cite{danzi2016anti}, is inspired by the concept of Software Defined Network (SDN)~\cite{ref17} that allows agile reconfiguration of the communication system due to \emph{function virtualization} and data-control separation. We therefore 
denote the proposed architecture as \emph{software-defined MicroGrid control}. 

In the proposed architecture, the data plane contains the actual information exchange between agents. The networking control plane is used by the agents to monitor and improve the links quality.
In our framework, the two planes are \emph{physically} separated via the use of different communication interfaces: wireless and power line communication (PLC), respectively.
As a PLC solution, we use power talk~\cite{ref3m}-\cite{ref5m}, a modem-less low-rate signalling technique designed specifically for the information exchange among controllers in a DC MG~\cite{stefanovic2017resilient}.
In this work, we enhance the operation of power talk with a Carrier Sense Multiple Access (CSMA) protocol, optimized for the use over this specific channel.
We also propose an algorithm for the dynamic selection of the active DERs (i.e. agents), which is run in a distributed manner by the PECs, relying on the information exchanged via power talk.
We show the benefits of the dynamic agent selection, investigating the scenarios in which there are dynamic changes in: (i) the wireless data plane connectivity among DERs; and (ii) the generated power within the MG.

The rest of the paper is organized as follows.
Section~\ref{sec:motiv} presents the background and motivation for this work.
Section~\ref{sec:model} introduces the system model.
Section~\ref{sec:pt} briefly presents power talk communications.
Section~\ref{sec:architecture} contains the description of the proposed architecture.
Section~\ref{sec:results} verifies the viability of the framework through a simulation implemented in Simulink.
Section~\ref{sec:conclusion} concludes the paper.

\section{Background and Motivation}\label{sec:motiv}

A number of robust networked control systems for DC MGs have emerged recently~\cite{wang2016improved}.
Besides the standard, centralized approach that is inherited from large-scale power systems, a distributed control implemented by multiple agents is a viable solution that mitigates the single-point-of-failure problem and increases the system modularity~\cite{meng2017review}.
The distributed control is typically based on consensus algorithm, which permits the agents to converge to a global average using only the information exchanges that are performed among, in communication terms, \emph{neighboring} agents~\cite{cady2015distributed,ref02,morstyn2016unified}. 
Relying on the broadcasting nature of the wireless medium, it was observed that wireless solutions overcome standard wired solutions in the support of gossip consensus algorithms, cf.~\cite{ref02}. 
Potential wireless solutions to be used in MGs are cellular networks~\cite{liang2013stability}, ZigBee~\cite{cady2015distributed}, Wi-Fi~\cite{ref02}, or their combinations~\cite{ref7m}, where the choice depends on the MG extent, affordable cost, and required Quality of Service.
However, the adoption of wireless technologies raises concerns regarding the reliability and security of the control.
In the following text, we present two motivating problems that our framework aims to solve.

\subsubsection{Selection of the active control agents}

\begin{figure}[!t]
\centering
{\includegraphics[width=0.7\columnwidth]{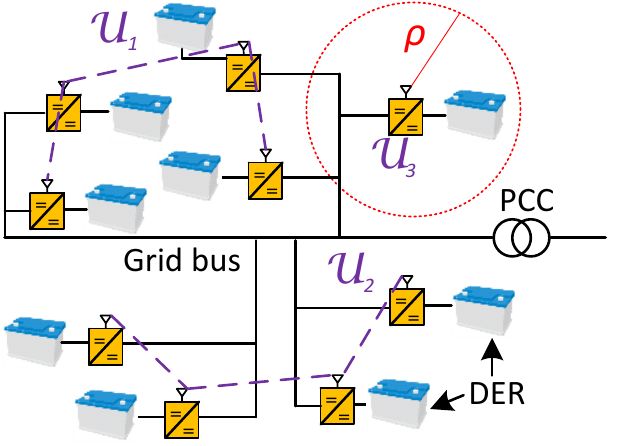}}
    \caption{Representation of a MG in which PECs are interconnected by electrical bus (black connections) and communication links (dashed purple).}
		\label{fig:subsets}
\end{figure}

In a power grid, the positioning of the wireless transceivers is constrained by the physical location of the DERs, such that certain agents may not be even able to establish communication to the rest of the agents.
This is illustrated in Fig.~\ref{fig:subsets}, where three communicating subsets are present, namely $\mathcal{U}_1$, $\mathcal{U}_2$ and $\mathcal{U}_3$.
Here $\mathcal{U}_3$ has a single isolated agent, that can neither communicate to other agents nor participate in the networked secondary/tertiary control.
Furthermore, a static selection of active agents, adopted in the state-of-art works, does not account for the \emph{cyber-physical} variations occurring in the MG, e.g., (i) variations of the power generated by DERs, (ii) varying reliability of the wireless channel and (iii) modifications of the communication network graph.
Therefore, the set of control agents should not be statically defined, but they should be able to adapt their control mode to the cyber-physical conditions of the MG.

\subsubsection{Cyber-security}

Employing off-the-shelf technologies, e.g. ZigBee or Wi-Fi, exposes the networked control to communication outages caused be external malicious entities, e.g. Denial Of Service (DoS) attacks.
Previous works~\cite{ref03} show that a malicious attacker equipped with a jamming device can easily undermine the information exchange, causing the system to converge to an erroneous voltage level.
Leveraging on the fact that the attacker is capable of jamming only a finite area, active DERs should be selected outside of its communication range, to be able exchange control information.

The key observation that underlies this work is that, besides the wireless network, DERs are interconnected by the electrical distribution network, which can be used as a side communication channel. This extra channel can bring the required resilience and security to the MG operation. 
The side channel is of a PLC nature; among the available PLC technologies, we adopt Power Line Signalling (PLS), which is a communication technique that does not require the installation of additional hardware, but relies on the PEC capabilities.
Specifically, we employ power talk~\cite{ref3m}-\cite{ref5m}, which can be implemented as a simple upgrade of the PECs' software.
As power talk provides a low-rate channel that is not capable of sustaining the secondary control data traffic\footnote{Note that the information required for the tertiary control can be exchanged via power talk~\cite{stefanovic2017resilient}.}, it is only used as a side, secure, and resilient communication channel to perform \emph{control-plane} exchanges that complement the wireless \emph{data-plane} communication links.

\section{System model}\label{sec:model}

The scenario considered in this paper is a single-bus DC MG in which a group of $U$ DERs is deployed to increase the self-sustainability of the installed load and to minimize the dependency to the main grid. 
We index the DERs in $\mathcal{U}=\left\{1,...,U\right\}$ and model the load as resistive impedance with power demand $P_d$.
DERs are interfaced to the distribution network bus by means of DC-DC PECs~\cite{ref2m}.
By $\mathbf{P}_M = [P_{1, M}, \cdots , P_{U, M}]$ we denote a vector that contains the maximum produced power, or capacity, of each DER.
We define the total generated power as $P_g = \sum_{u\in \mathcal{U}} P_{u, M}$.

In our model, PECs have the dual-mode capability~\cite{ref12}: their software implements the logic of both Current Source Control (CSC) and Voltage Source Control (VSC), and they are able to switch between these operation modes seamlessly.
This feature permits to use the interfaced DER at its full capacity when performing CSC, as well as to participate in the voltage control when performing VSC.
Also, a stand-alone grid, such as an islanded MG, requires that a subset of DERs $\mathcal{V}$, $\mathcal{V} \in \mathcal{U}$ and $|\mathcal{V}| = V> 0$, performs VSC.
This subset is called Voltage Source Set (VSS), where the set cardinality depends on the capacity of the resources.
The remaining $U-V$ DERs from the set $\mathcal{U} \setminus \mathcal{V}$ perform CSC, thus constituting the passive agents of the network.

\subsection{Voltage Source Control}

The VSC is organized according to a hierarchical architecture~\cite{ref01}, composed of three levels.

\subsubsection{Primary control}

A droop loop accomplishes the current sharing among the VSC units.
The loop has a sampling/control period $T^\text{pc}$ and can be expressed at DER $u$ as 
\begin{equation}\nonumber
v_u = v_\text{ref} - r_{u}i_u, \; u\in\mathcal{V},
\end{equation}
where $v_u$ and $i_u$ are the output voltage and current, $v_\text{ref}$ is the reference voltage and $r_{u}$ is the virtual resistance.
The loop is implemented by a PI controller with proportional parameter $K^\text{pc}_\text{p}$ and integral parameter $K^\text{pc}_\text{i}$.
This control loop depends only on local measurements, which causes the bus voltage deviations from the reference value $v_\text{ref}$.

\subsubsection{Secondary control}\label{sec:sec_con}

A distributed secondary control is in charge of (i) the restoration of the voltage to $v_\text{ref}$ and (ii) the proportional current sharing among generators~\cite{ref01}, and is based on the local estimation of the average voltage $\bar{v}_u$ and current $\bar{i}_u$ in the MG.
Among the multitude of distributed secondary control schemes~\cite{wang2016improved}, we adopt the solution proposed in~\cite{meng2016modeling}, which utilizes two compensation terms $\delta x^\text{i}$ and $\delta x^{\text{v}}$ to balance the equation
\begin{equation}\label{eq:v_star}
	v_u^{\star} = v_\text{ref} + \delta x^{\text{i}} + \delta x^{\text{v}} - r_u i_{u},\;u\in\mathcal{V}.
\end{equation}
Two PI controllers, characterized by proportional and integral parameters $K^\text{sc}_{v,\text{p}}$, $K^\text{sc}_{v,\text{i}}$ and $K^\text{sc}_{i,\text{p}}$, $K^\text{sc}_{i,\text{i}}$, respectively, generate the correction terms according to:
\begin{equation}\label{eq:correction}
	\delta x^{\text{i}} = \tilde{i}_{u} - \alpha_u \cdot V \cdot \bar{i}_u, \qquad \delta x^{\text{v}} = \tilde{v}_{u} - \bar{v}_u,
\end{equation}
where
\begin{equation}\label{eq:alpha}
\alpha_u = \frac{P_{u, M} }{\sum_{v=0}^{V}P_{v, M}}.
\end{equation}
The scheme requires the global knowledge of $\mathbf{P}_{M}$, but improves the one adopted in~\cite{danzi2016anti} by promoting the proportional power sharing among VSC DERs.
We compute the average values through a \emph{robust broadcast gossiping}~\cite{ref02}, suitably modified to include the proportional term $\alpha_u$.
The algorithm requires the synchronization among agents, being executed in periods of duration $T^{\text{sc}}$, which also determine the sampling frequency of the secondary control.
The secondary control operation iterates through the following steps: 
\begin{enumerate}
\item In secondary control period $k$, DER~$u\in\mathcal{V}$ broadcasts a packet to the neighboring control agents over the wireless interface with payload $\mathbf{a}_u(k) = [\bar{v}_{u}(k), \bar{i}_{u}(k)]$, representing the local estimates of the average voltage and current measured by DERs, respectively.
\item During the same secondary control period $k$, DER~$u$ receives $R$ packets broadcasted from neighboring agents, which are supplied as input to the robust broadcast gossip algorithm:
\begin{equation}\nonumber
\mathbf{a}_u(k+1) = \left\{
  \begin{array}{lr}
    \beta_u \mathbf{m}_{u}(k) + ( 1 - \beta_u ) \frac{\sum_{j}\mathbf{a}_{j} (k)}{R}, & R>0\\
    \beta_u \mathbf{m}_{u}(k) + ( 1 - \beta_u ) \mathbf{a}_{u}^{'} (k), & R=0
  \end{array}
\right.
\end{equation}
where $\mathbf{m}_{u}(k)=[\tilde{v}_u(k),\tilde{i}_u(k)]$ is the local measurement vector in secondary control period $k$, $\mathbf{a}_{j} (k)$ is the status of agent $j$ (assumed to be within the communication range of agent $u$), $\mathbf{a}_{u}^{'}(k) = [0, \tilde{i}_u(k)/(\alpha_u \cdot V)]$ serves to compensate \eqref{eq:alpha} when information from neighbors is not available and $\beta_u\in (0, 1)$ is the consensus weight for agent $u$; in this paper we adopt that $\beta_u=V^{-1}, \; \forall u\in\mathcal{V} $.
\item The new status $\mathbf{a}_u(k+1)$ is then forwarded to the local PI controllers, together with the measurement vector $\mathbf{m}_u(k)$.
\item The PI controllers generate the reference voltage correction signals $\delta x^{\text{i}}$ and $\delta x^{\text{v}}$ of \eqref{eq:correction}.
\end{enumerate}

In case of a load/generation change, the offsets $\delta x^{\text{i}}$ and $\delta x^{\text{v}}$ converge to stable values when a global consensus is reached by the described scheme.
Note that the adopted broadcast gossip consensus algorithm provides high robustness against packet losses, see \cite{ref02}, because the local information is used in absence of external one.

\subsubsection{Tertiary control} The tertiary control is in charge of the economic optimization, as well as the management of the bidirectional power flow with the main grid. 
The information exchange happens on lager time intervals, typically $T^\text{tc}=5-30$ minutes. 
For this reason, the communication requirements are less demanding and, in this work, we don't adopt a tertiary control, but the proposed model can be easily extended.
We further assume that $\mathbf{P}_M$ is constant over a tertiary control period and that the demand is always satisfied, i.e. $P_g > P_d$.

\subsection{Current Source Control}

A PEC~$u$ that performs CSC permits the utilization of the DER as a passive agent.
The electrical model consists of a current generator $i_{u, M}$ in parallel with an internal resistance approximated by $r_d$~\cite{ref12}.
As opposed to VSC DERs, CSC DERs do not participate in the active control and are excluded from the broadcast gossip algorithm.

\subsection{Communication System}

Each DER is equipped with an IEEE 802.11 wireless interface that supports the ad-hoc network mode, allowing decentralized information exchange and multicast packet transmission.
The wireless interface is characterized by the communication range $\rho$, see Fig.~\ref{fig:subsets}, which determines the set of neighbors reachable by DER~$u$, indicated with $\mathcal{N}_u$.
We assume that the links are bidirectional, i.e. the communication graph is undirected.
Depending on the DERs positions and $\rho$, the wireless network graph $\mathcal{G}$ may be split into $\omega \geq 1$ subgraphs, where the $j$-th graph is denoted by $\mathcal{G}_{j}$, its set of DERs by $\mathcal{U}_j$, and $\cup_{1 \leq j \leq \omega}\mathcal{U}_{j}=\mathcal{U}$.
To ensure convergence of the consensus algorithm, the graph connecting $\mathcal{V}$ should be strongly connected, i.e., $\mathcal{V} \subseteq \mathcal{U}_j$ for some $j$.
The wireless network traffic consists of packets containing information required to run the secondary control, which are generated by DER $u\in\mathcal{V}$.
Table~\ref{table:comparison} gives a comparison between existing distributed secondary control schemes.
Note that the number of DERs is usually limited, and the secondary control traffic has a duty cycle in the order of tens of milliseconds.

\begin{table}[]
\centering
\caption{Comparison of existing communication models for distributed secondary control; n.d. means \emph{not declared}.}
\begin{tabularx}{\columnwidth}{l|l|l|l|l}
\hline
\textbf{Paper} & \multicolumn{2}{l|}{\textbf{Traffic}} & \multicolumn{2}{l}{\textbf{Network}}    \\ \hline
               & \textbf{$T^\text{sc}$ [ms]} & \textbf{Packet size {[}bit{]}} & \textbf{$U$} & \textbf{Graph}       \\ \hline
\cite{meng2016modeling}& 100-1000 & n.d.  & 6  & Mesh, cross, ring, line \\
\cite{ref02}           & 10      & n.d.  & 4  & Mesh     \\
\cite{ref03} & 25 & 16  & 6 & Mesh  \\
\cite{cady2015distributed} & 50 & $\sim 800$ & 3 & Mesh, directed \\  
\cite{danzi2016anti} & 25 & 16 & 9 & Random    \\
\cite{morstyn2016unified} & 20 & n.d. & 10 & Mesh           
\end{tabularx}
\label{table:comparison}
\end{table}

\subsection{Metrics}
We use $i_{v}^{*}$ to denote the output current of VSC DER~$v \in\mathcal{V}$, in the ideal case when the grid voltage is $v_\text{ref}$.
The sub-optimality of the secondary control with respect to the ideal case is expressed as the sum of the deviations of the output current of VSC DERs with respect to the corresponding $i_{v}^{*}$:
\begin{equation}
J^\text{sc} = \sum_{v \in \mathcal{V}} |i_{v} - i_{v}^{*}|.
\end{equation}

We define $J^\text{cq}$ as the \emph{quality} of specific selection of VSC set.
In general, $J^\text{cq}$ can depend both on physical properties, such as the generation capacity, as well as cyber aspects, i.e. the wireless network graph connectivity or communication cost.
We define $J^\text{cq}$ as the probability that the power demand is lower than the CSC DERs' production, i.e. the probability that VSC DERs are absorbing energy:
\begin{equation}
J^\text{cq} = \mathrm{Prob}\left\{ P_d < \sum_{u \in \mathcal{U} \textbackslash \mathcal{V}} P_u \right\}.
\end{equation}
The above definition requires knowledge of the demand statistics, which can be modeled as Gaussian~\cite{xia2016probabilistic}, i.e. $P_d \sim \mathcal{N}(\mu, \sigma^2)$. Then, we have:
\begin{equation}
J^\text{cq} = 1- Q( x ), \; x = \frac{\sum_{u \in \mathcal{U} \textbackslash \mathcal{V}} P_u - \mu}{\sigma}
\end{equation}
where $Q(x)$ is the Q-function.

\subsection{Selection of Voltage Source Controllers}\label{sec:consensus}

\begin{figure}[!t]
\centering
{\includegraphics[width=0.7\columnwidth]{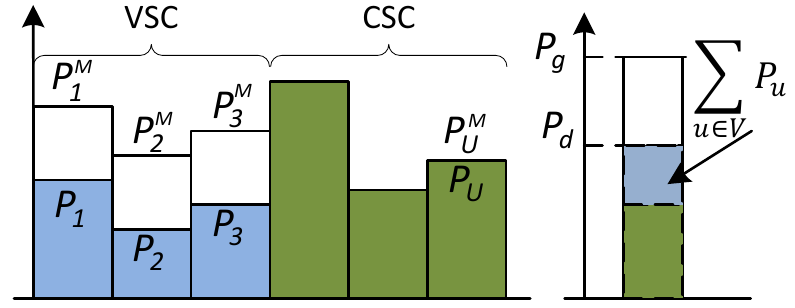}}
    \caption{Representation of the power sharing among DERs.}
		\label{fig:capacity}
\end{figure}

Different choices of DERs in $\mathcal{V}$ lead to different $J^\text{sc}$ and $J^\text{cq}$, making necessary the formulation of a proper selection algorithm for the VSC set.
As illustrated in Fig.~\ref{fig:capacity}, the set cardinality, $V$, should allow to meet the variations of demand: if too small, the VSC DERs may have to absorb energy; on the other hand, the inclusion of all DERs in $\mathcal{V}$ increases the communication cost and reduces the DER utilization efficiency.
Furthermore, as discussed in Section~\ref{sec:motiv}, possible choices of VSCs are constrained by the cyber network topology, i.e. all VSCs should be in the same, strongly connected, subset. 
The objective that we set is to find the \emph{connected} set of DERs with \emph{lowest cardinality} that gives
\begin{equation}\label{eq:requirement}
J^\text{cq} < p_\text{abs},
\end{equation}
where $0 \leq p_\text{abs} \leq 1$ is the highest allowed probability of the event in which VSC DERs have to absorb power.
If there is no solution to \eqref{eq:requirement}, we select the connected graph that provides the lowest $J^\text{cq}$, to ensure that the algorithm always gives solution.
Leveraging on the fact that in a typical scenario the set of DERs has low cardinality, see Table~\ref{table:comparison}, we propose Alg.~1.
This is an exhaustive algorithm that builds an ordered collection $\mathbfcal{L}$ of all connected graphs, then selects the one that satisfies the selection conditions.
The connectivity of graph connecting a generic set of DERs, $\mathcal{W}$, can be verified, for instance, by means of the well-known Tarjan's algorithm~\cite{tarjan1972depth}.
\begin{algorithm}
\caption{VSC set selector}\label{alg:vscset}
\begin{algorithmic}
\State input $\mathbf{P_M}$, $\mathcal{G}_{\mathcal{U}}$
\State $\mathcal{V} \gets \emptyset$, $\mathcal{V'} \gets \emptyset$, $\hat{p} \gets 1$, $\mathbfcal{L} \gets \bigcup\limits_{ \{ \mathcal{W} : \mathcal{W} \in \mathcal{U}, |\mathcal{W}| \leq V \}} \mathcal{W}$
\For {$\mathcal{W}$ in $\mathbfcal{L}$}
\If {$\mathcal{W}$ is not connected}
	\State $\mathbfcal{L} \gets \mathbfcal{L} \textbackslash \mathcal{W}$
\Else
	\State $p \gets J^\text{cq}(\mathbf{P_M}, \mathcal{G}_{\mathcal{W}})$
	\If {$p \leq \hat{p}$}
		\State $\hat{p} \gets p$, $\mathcal{V'} \gets \mathcal{W}$
	\EndIf
	\If {$p \geq p_\text{abs}$}
		\State $\mathbfcal{L} \gets \mathbfcal{L} \textbackslash \mathcal{W}$
	\EndIf
\EndIf
\EndFor
\If {$|\mathbfcal{L}| > 0$}	
	\State $\mathcal{V} \gets \mathbfcal{L}[0]$
\EndIf
\State output $\mathcal{V}$
\end{algorithmic}
\end{algorithm}

\section{Power Talk}\label{sec:pt}

Power talk is a communication technique implemented within the primary control loops of the PECs
~\cite{stefanovic2017resilient}.
The principle idea is to send information over the power grid buses by varying the droop control parameters of VSC DERs; in this paper, we focus on the scenario where the information is conveyed only through the variations of the reference voltage $v_\text{ref}$.
These variations cause the variations of the steady state voltage that is observed by other PECs.
We note that, in order to deviate the steady state voltage, the secondary control has to be temporary suspended in the grid during the signalling periods of power talk.
In principle, power talk can use any modulation scheme \cite{ref3m} and channel coding.
In this paper, we adopt a binary modulation, with symbol period $T^\text{pt}$, amplitude $a^\text{pt}$, and no channel coding, cf.~\cite{ref3m}-\cite{ref5m}.
The observations of the transmitted symbols periods are with a lag of $\tau$ seconds, in order allow the bus to reach a steady state after $v_\text{ref}$ has been deviated~\cite{ref4m}.

\subsection{Power-Talk Communication Protocol}
\label{sec:ptchannel}

\begin{figure}[t]
\centering
{\includegraphics[width=\columnwidth]{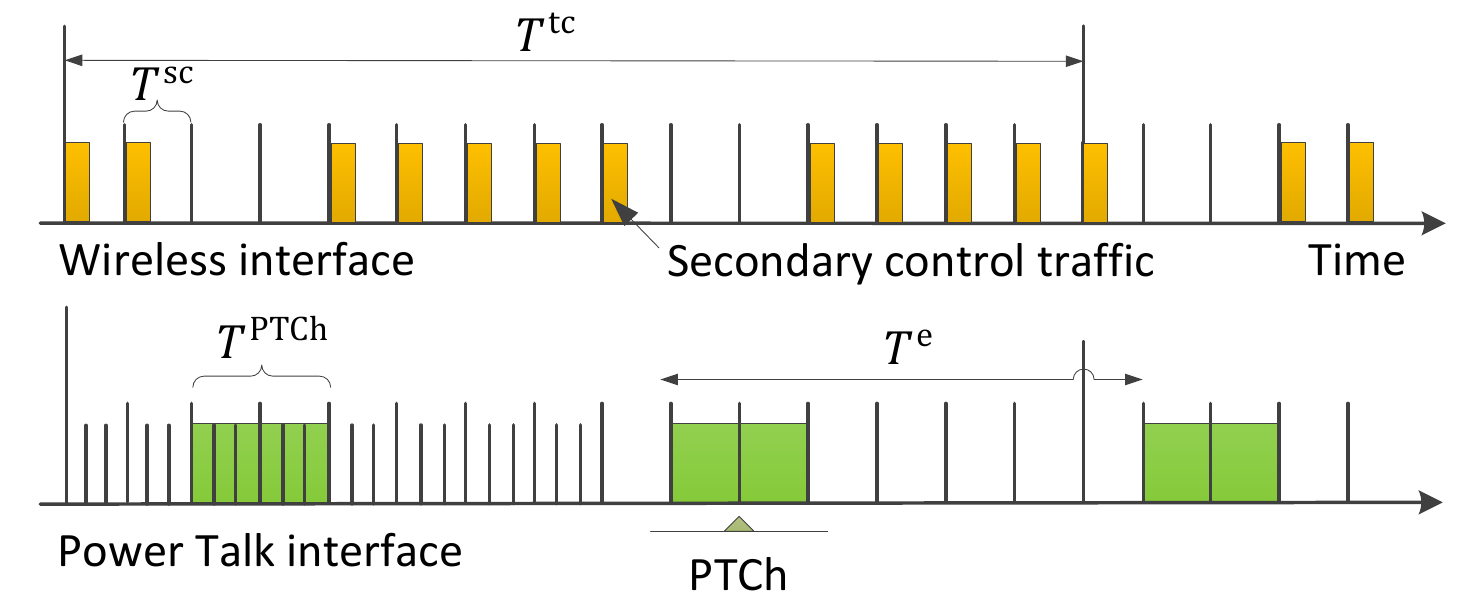}}
    \caption{Time relationship between power talk and wireless interfaces.}
		\label{fig:timeax}
\end{figure}

We assume that signalling via power talk is performed periodically, through a power talk channel (PTCh) that is established in intervals of duration $T^{\text{PTCh}}$ seconds.
As already noted, the secondary control is suspended during the PTCh, and, thus, $T^\text{PTCh}$ should be as short as possible.
The PTCh's are triggered with a constant period: this permits to coordinate the VSC DERs in the suspension of the secondary control and keeps the design complexity low.
In order to enable plug-and-play capability of the MG, here we enhance power talk with a Carrier Sense Multiple Access (CSMA) scheme. This enables random access to the shared channel, while keeping the time in which the communication system is idle low.
Specifically, a CSMA protocol coordinates DER transmissions during $T^\text{PTCh}$.
The protocol is characterized by the carrier sensing duration, or virtual slot, $\sigma_\text{s}$, that is chosen based on the propagation delay in the grid.
The channel access operates as follows.
At the beginning of each PTCh period, each DER sets an internal counter to a uniformly randomly selected number in $[0, B-1]$, where $B \in \mathbb{N}$ is a design parameter.
The counter is decremented for each idle virtual slot, $\sigma_\text{s}$, and, when it reaches zero, the DER starts the transmission.
When a DER starts transmitting, the others stop decrementing their counter for the duration of a transmission, $T^\text{dt}$, that is known a priori.
In case of a collision, when more than one DER selects the same number of slots, the corresponding DERs do not retransmit.
Instead, they keep the collision probability in a PTCh to be below a low target value through the appropriate selection of $B$, as described in the following section.

\section{The Software-Defined Control Architecture}\label{sec:architecture}
To demonstrate the advantages of the proposed software-defined control architecture, in the following text we present the potential application of decentralized voltage sources set selection.

\subsection{Decentralized Voltage Sources Set Selection (DVSSS)}

The operation of MGs is subject to variations in the generated power, wireless connectivity among the agents, due to e.g. wireless jamming from malicious agents, and grid topology, such as addition/removal of DERs and loads.
Thus, Alg.~1 should be run periodically, e.g., every tertiary control period, in order to determine the current optimal set of VSC DERs $\mathcal{V}$. Due to the lack of centralized coordination, this should be done in a decentralized manner: DERs receive updated information about DERs' power capacity and connectivity, $\mathbf{P_M}$ and $\mathcal{G}$, execute Alg.~1 and thereby set their control mode.

\begin{figure}[t]
\centering
{\includegraphics[width=0.9\columnwidth]{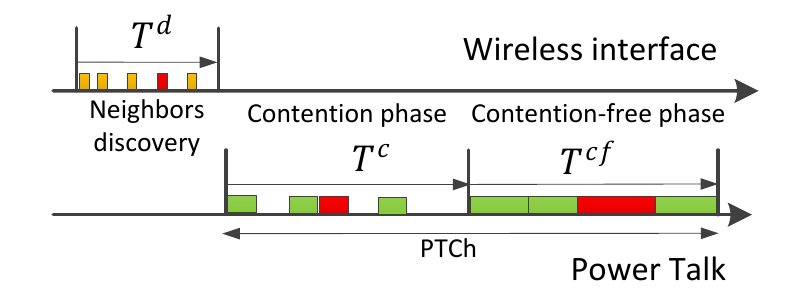}}
    \caption{Wireless and power talk channel as seen by DER $u$ during DVSSS messages exchange. Information transmitted by $u$ is coloured in red.}
		\label{fig:vsss}
\end{figure}

The scheme, indicated as decentralized voltage sources set selection (DVSSS), is executed periodically and consists of the following steps, represented in Fig.~\ref{fig:vsss}:
\begin{enumerate}
\item All DERs activate the wireless interface and broadcast short, control plane packets for $T^\text{d}$ seconds. This step is required to notify their presence and obtain the list of their potential neighbors (\emph{neighbors discovery phase}), and is executed $T^\text{d}$ seconds before the beginning of the power talk channel (PTCh).
\item DERs broadcast the information about their neighbors and generation capacity over the power talk interface.
\item Each of them locally executes Alg.~1.
\item If DER $u\in \mathcal{V}$, then activate the VSC, tuning the consensus algorithm according to \eqref{eq:alpha}; otherwise activate the CSC and do not transmit over the wireless channel until the next subset selection.
\end{enumerate}

For simplicity, the DVSSS is run in every tertiary control period, $T^\text{tc}$. The PTCh, used for control plane exchanges, is established with the same periodicity, i.e., every $T^\text{e} = T^\text{tc}$ seconds, and its duration, $T^\text{PTCh}$ seconds, is divided in two phases.
The first one is a CSMA-based contention phase of duration $T^\text{c}$ seconds, in which DERs \emph{notify their presence}, operation required to permit addition/removal of DERs.
The second phase is contention-free and of duration $T^\text{cf}$ seconds, in which each DER $u$ broadcasts the lists of its wireless neighbours $\mathcal{N}_u$. The total duration is:
\begin{equation}
T^\text{PTCh} = T^\text{c} + T^\text{cf}.
\end{equation}
During $T^\text{c}$, each DER $u$ contends the channel by means of the described CSMA protocol, and broadcasts a $b-$bit packet that contains its own ID and $P_{u,M}$.
The duration of $T^\text{c}$ depends on (i) the CSMA parameters $\sigma_\text{s}$ and $B$, (ii) $p_\text{c}$, defined as the probability that all $U$ DERs choose a different counter value $B$ and there are no collisions, and (iii) the probability that all agents attempt the transmission before $T^\text{c}$.
We require that $p_\text{c} > p^{\star}_\text{c}$, where $p^{\star}_\text{c}$ is a target probability, and all DERs' counters value reach zero before $T^\text{c}$, which certainly happens if the number of idle virtual slots, $N_b$, is greater than $B$.
These two conditions are, respectively:
\begin{equation}\label{eq:pc}
p_\text{c} = \frac{B!/(B-U)!}{B^U} > p^{\star}_\text{c}, \; N_b = \left\lceil{\frac{ T^\text{c} - U\cdot T^\text{dt} }{\sigma_\text{s}}}\right\rceil \geq B,
\end{equation}
where  $T^\text{dt} = b \cdot T^\text{pt} + \tau$ is the duration of the transmission of one DER and $\tau$ is a guard time to account for the network propagation delay.
If we use $U$ from the previous period, and fix a target $p^{\star}_\text{c}$, we can find $B$ and $T_\text{c}$.

At the end of the contention period, DERs have the information about $U$ as well as an ordered list of DERs' IDs.
In the contention-free phase, each DER access the channel in a token-based manner, where the token passing follows the list of IDs.
The duration of this phase is simply
\begin{equation}\label{eq:cf}
T^\text{cf} = b \cdot T^\text{pt} \cdot e + U\cdot \tau,
\end{equation}
where $e$ is the number of edges of the network graph, while $U\cdot \tau$ stands for the addition of one guard interval per DER.

\section{Results}\label{sec:results}

\begin{figure}[!tb]
\centering
\subfloat[]{\includegraphics[width=0.75\columnwidth]{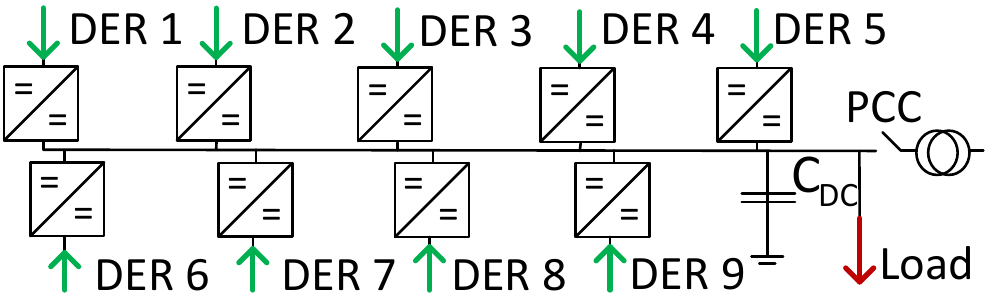}}

\subfloat[]{\includegraphics[width=\columnwidth]{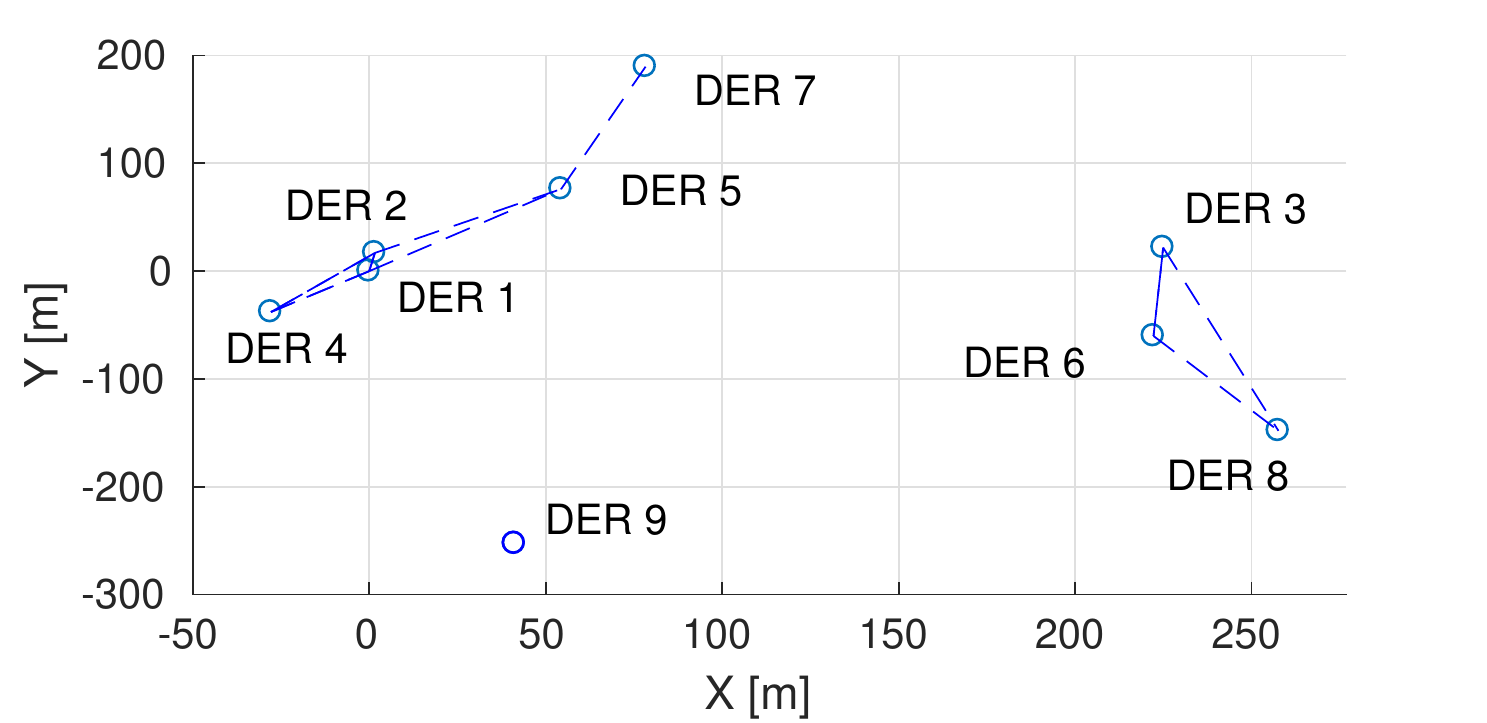}}
    \caption{The case study MG. (a) Electrical scheme of the power talk communication network. The arrow up represents a load, the arrow down a DER. (b) Spatial position of DERs. The dashed lines are wireless communication links.}
 \label{fig:case_study}
\end{figure}
\begin{table}[]
\centering
\caption{Simulation parameters.}
\label{tab:simulation}
\begin{tabular}{llllll}
\hline
\multicolumn{6}{l}{Communication and algorithm parameters} \\ \hline
$\rho$     & 175 m & $T^\text{d}$ & 10 s &    $p_\text{abs}$      & 0.01 \\ \hline
\multicolumn{6}{l}{Power talk parameters} \\ \hline
$\sigma_\text{s}$ & 0.003 s     & $b$                     & 8 bits    & $T^\text{pt}$ & 0.01 s \\ 
$a^\text{pt}$ & 0.5 V &  &  & & \\ \hline
\multicolumn{6}{l}{Control parameters}                        \\ \hline
$T^\text{pc}$ & $10^{-4}$ s & $T^\text{sc}$ & $10^{-2}$ s & $r_\text{d}$ & 0.385 \\ 
$K^\text{sc}_{v,p}$ & 0.1 & $K^\text{sc}_{v,i}$ & 20 & $K^\text{sc}_{i,p}$ & 0.1  \\ 
$K^\text{sc}_{i,i}$ & 20 & $K^\text{pc}_{p}$ & 3 & $K^\text{pc}_{i}$ & 170 \\ \hline
\multicolumn{6}{l}{Electrical parameters}                        \\ \hline
$v_\text{ref}$ & 380 V & $C_\text{DC}$ & $2.2 \cdot 10^{-3}$ F & $\mu$ & 14.44 kW \\
$\sigma^2$ & 1.44 kW & $p_{\mathrm{min}}$ & 1 kW & $p_{\mathrm{max}}$ & 4 kW \\ \hline
\end{tabular}
\end{table}
The proposed framework is verified through numerical simulation of a case-study DC MG with $v_\text{ref} = 380$~V and $U=9$ DERs, which is isolated from the main grid at the point of common coupling (PCC).
The overall electrical system, represented in Fig.~\ref{fig:case_study}(a), is inspired by~\cite{morstyn2016unified}.
Here we also take into account the relative positions of the DERs, see Fig.~\ref{fig:case_study}(b), where a wireless communication link (depicted in blue) between two DERs exists only if they are in the communication range $\rho$ of each other.
Further, the capacity $C_\text{DC}$ models the cumulative capacities of DERs' output filters.
The MG, parametrized as in Table~\ref{tab:simulation}, is implemented in a MATLAB/Simulink simulator.

\subsubsection{PTCh duration}

The propagation delay observed in the grid is $\tau = 2.8$~ms, and is used to set the value of $\sigma_\text{s}$.
In the next step, we fix $p^{\star}_c = 0.8$ and solve \eqref{eq:pc}, obtaining $B=165$ and $T^\text{c} = 1.24$~s.
For $U=9$, a full-mesh graph has $e = 36$ edges, and we use these values to solve \eqref{eq:cf} and obtain $T^\text{cf} = 2.91$.
Finally, we set the duration of PTCh in our simulation as $T^\text{PTCh} = T^\text{c} + T^\text{cf} = 1.24 + 2.91 = 4.15$~s.

\subsubsection{Static Vs Dynamic set selection in case of DoS attacks}

\begin{figure}[!tb]
\centering
\subfloat[]{\includegraphics[width=0.78\columnwidth]{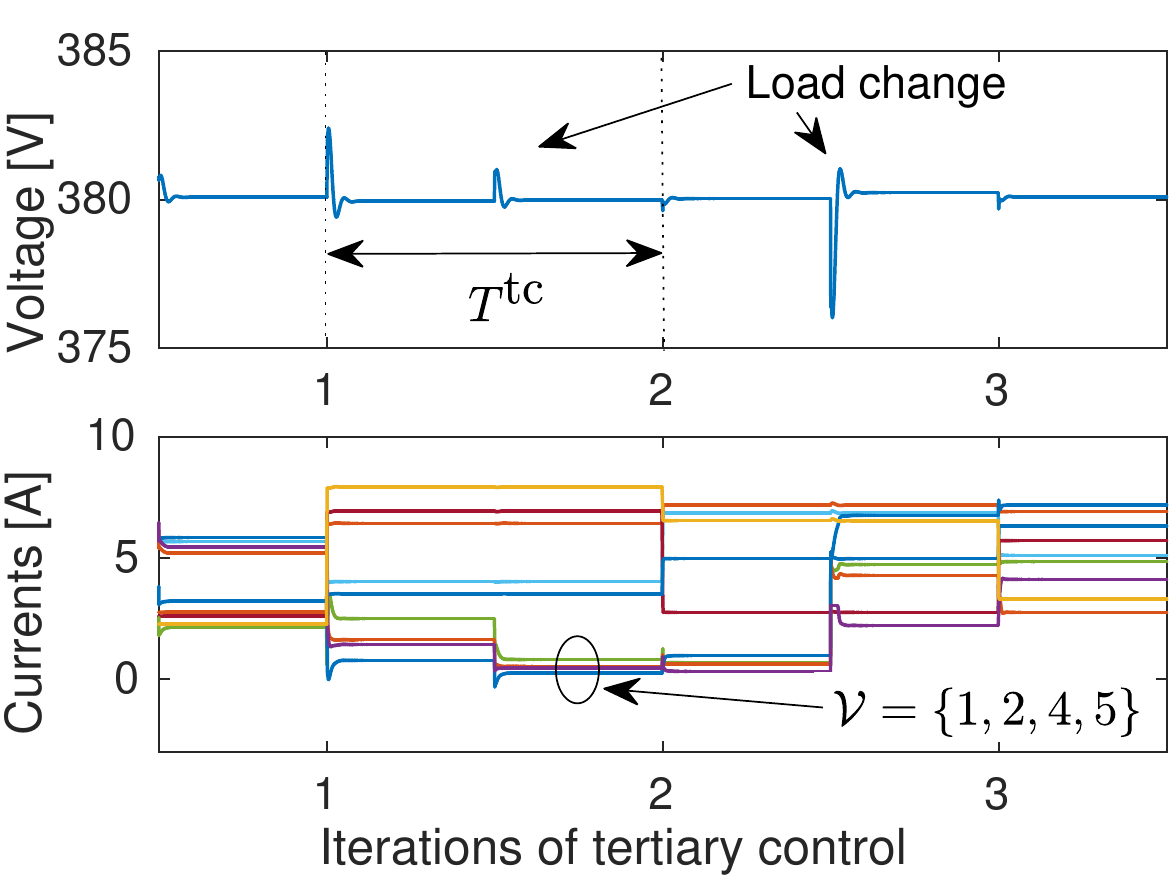}}

\subfloat[]{\includegraphics[width=0.78\columnwidth]{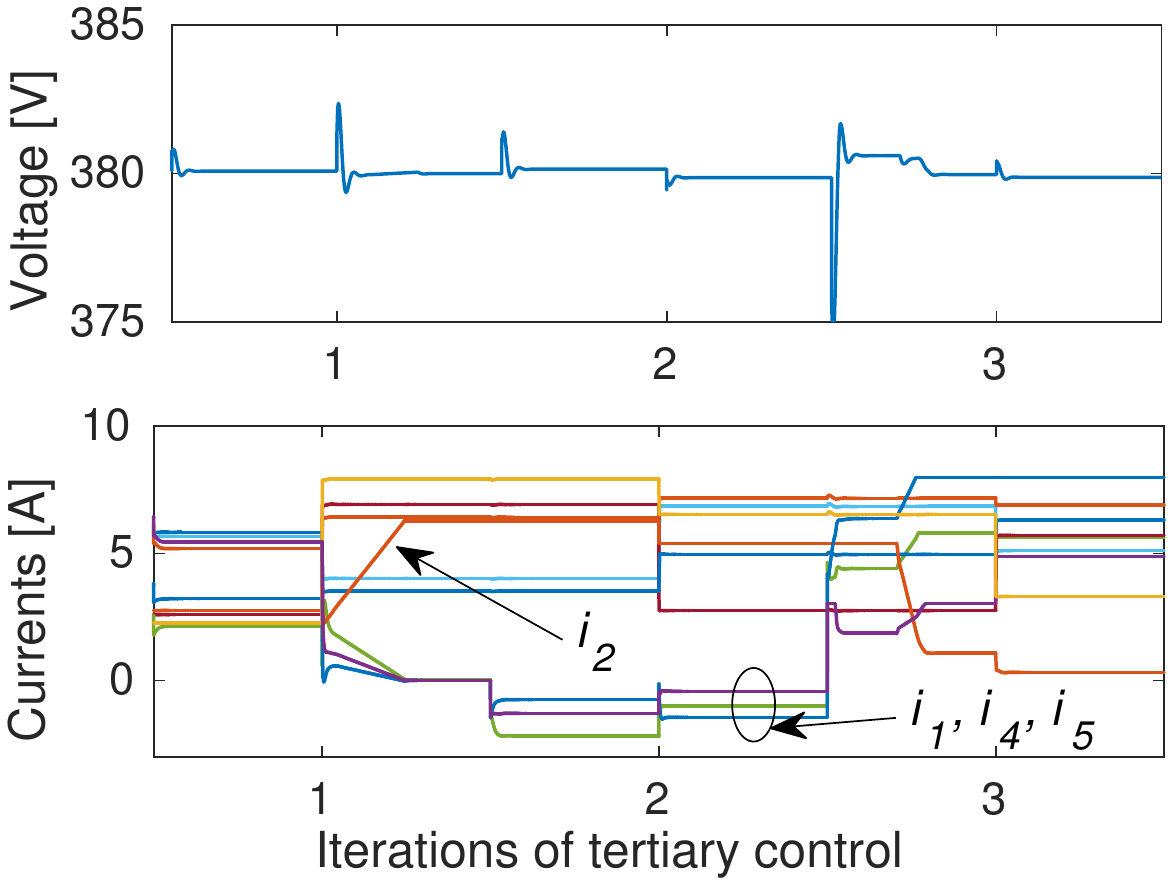}}
    \caption{Realization of the process when the wireless network graph $\mathcal{G}$ is (a) stable and (b) split due to a DoS attack.}
 \label{fig:currents_exp}
\end{figure}
\begin{figure}[!tb]
\centering
{\includegraphics[width=\columnwidth]{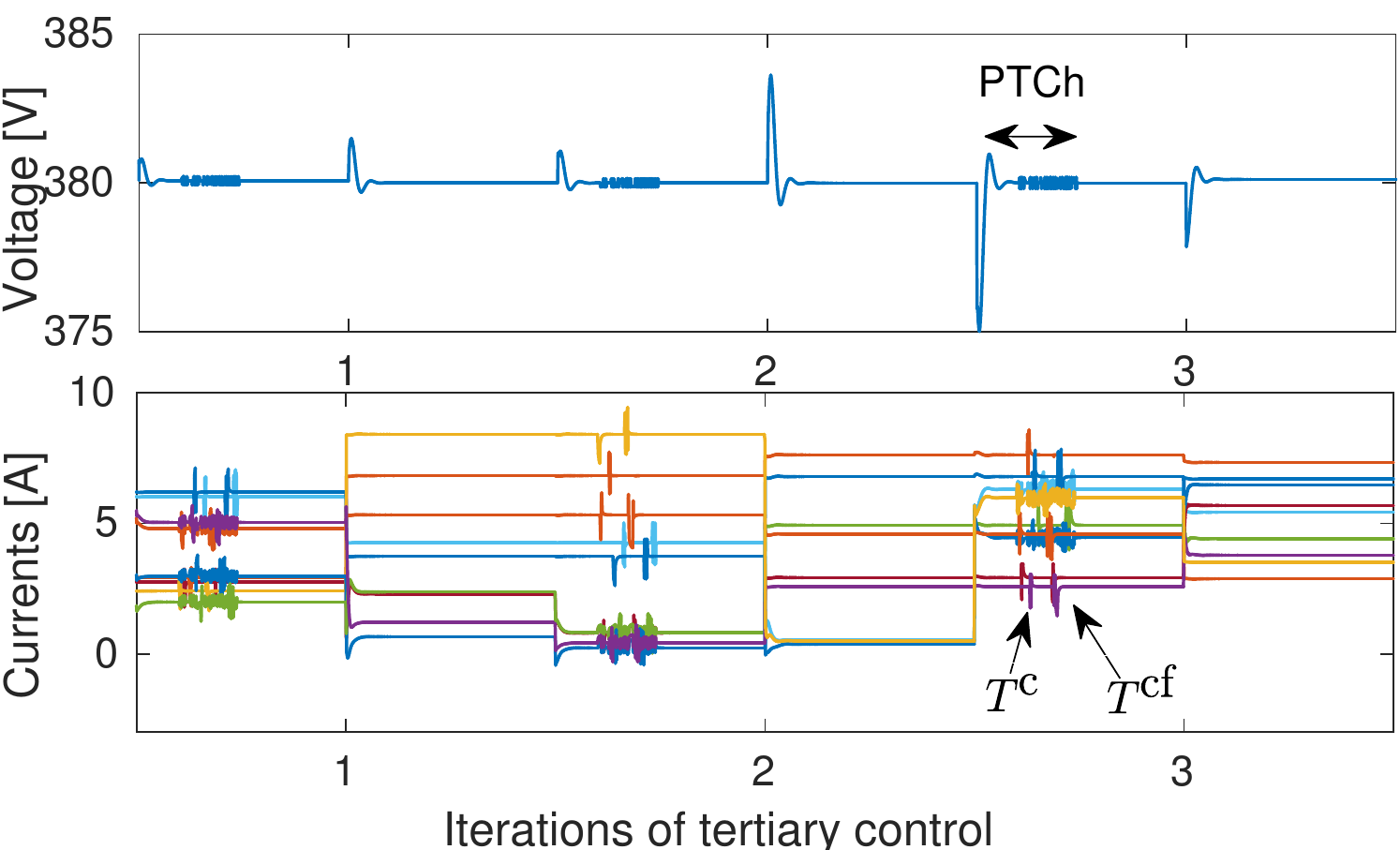}}
    \caption{Realization when the DVSSSs selection is applied to counteract DoS.}
 \label{fig:case_solution}
\end{figure}

We initially assume that set of VSC DERs is $\mathcal{V} = \{1, 2, 4, 5\}$, that form a strongly connected graph. 
In each tertiary control period, the capacity of DERs is selected uniformly in $[ p_{\mathrm{min}}, p_{\mathrm{max}}]$, with values reported in Table~\ref{tab:simulation}.
In Fig.~\ref{fig:currents_exp}, we report voltage measured at PCC and currents measured by DERs, during a simulation with duration $3 T^\text{tc}$.
In the middle of each tertiary control period, a random load variation is triggered, to show the system adaptation to the load change. 
Initially, the VSC set is static and connected, see Fig.~\ref{fig:currents_exp}(a), the power injection is proportionally shared among DERs and the voltage is kept at the reference value $v_\text{ref}$.
Fig.~\ref{fig:currents_exp}(b) shows the case when communication of DER 2 is jammed by an external attacker and VSC communication graph is split in two sub-sets $\{1, 4, 5\}$ and $\{2\}$, resulting in unbalanced power sharing.
In particular, observe that DER~2 becomes isolated and prevented to participate to the consensus, injecting more current than required and causing the other three VSC DERs to compensate the error.
We report in Table~\ref{table:metrics} the value of the metrics that express the secondary control sub-optimality and quality, respectively $J^\text{sc}$ ampere and $J^\text{cq}$ at the beginning of a tertiary control period, when the steady state is reached. Table~\ref{table:metrics}(a), that corresponds to the simulation of Fig.~\ref{fig:currents_exp}(a), shows that the secondary control cost is $J^\text{sc} = 0$ in each period, since $i_{v} = i_{v}^{\star}, \; \forall v \in \mathcal{V}$. Table~\ref{table:metrics}(b) corresponds to the simulation of Fig.~\ref{fig:currents_exp}(b), and shows how the graph split reflects in a positive $J^\text{sc}$.
The results clearly motivate the need of a dynamic selection algorithm for the DERs' operating mode. 

Fig.~\ref{fig:case_solution} reports the simulation of the same process when we apply the proposed DVSSS algorithm. 
The VSC set, $\mathcal{V} = \{1,2,4,5\}$, is split by the attacker, but the information exchange over PTCh permits the execution of DVSSS algorithm and the establishment of a new connected set $\mathcal{V} = \{1,4,5,7\}$, excluding DER~2 from the active controllers and restoring the wireless network connectivity.
In the second iteration of the DVSSS, a new set $\mathcal{V} = \{3,6,8\}$ is established, since it provides the lowest $J^\text{cq}$ in that tertiary control period.
Compared to Fig.~\ref{fig:currents_exp}(b), the algorithm reduces the cost $J^\text{sc}$ to zero in each iteration as well as decreases $J^\text{cq}$ to $1.7\times 10^{-4}$, $5.7\times 10^{-4}$, $1.9\times 10^{-6}$.
The figure also shows a sample of the period when power talk takes place, which includes a CSMA phase of duration $T^\text{c}$ and a contention-free phase of duration $T^\text{cf}$.

\begin{table}[]
\centering
\caption{Comparison of metrics. (a) corresponds to the simulation of Fig.~\ref{fig:currents_exp} and (b) to the one of Fig.~\ref{fig:case_solution}.}
\label{table:metrics}
\subfloat[]{
\begin{tabular}[width=0.3\columwidth]{lll}
\hline
\multicolumn{1}{l}{Iteration} & \multicolumn{1}{l}{$J^{cq}$} & \multicolumn{1}{l}{$J^{sc}$ [A]} \\ \hline
1                               & $2.5\times 10^{-3}$                         & 0                             \\
2                               & $1.3\times 10^{-3}$                         & 0                             \\ 
3                               & $3.3\times 10^{-4}$                         & 0                           \\ \hline
\end{tabular}}
\hfill
\subfloat[]{
\begin{tabular}[width=0.3\columwidth]{lll}
\hline
\multicolumn{1}{l}{Iter.} & \multicolumn{1}{l}{$J^{cq}$} & \multicolumn{1}{l}{$J^{sc}$ [A]} \\ \hline
1                               & $2.5\times 10^{-3}$                         & 9.28                             \\ 
2                               & $1.3\times 10^{-3}$                        & 9.60                             \\ 
3                               & $3.3\times 10^{-4}$                         & 4.80                            \\ \hline
\end{tabular}
}
\end{table}

\subsubsection{DVSSS and control quality improvement}

\begin{figure}[!tb]
\centering

\subfloat[]{\includegraphics[width=0.85\columnwidth]{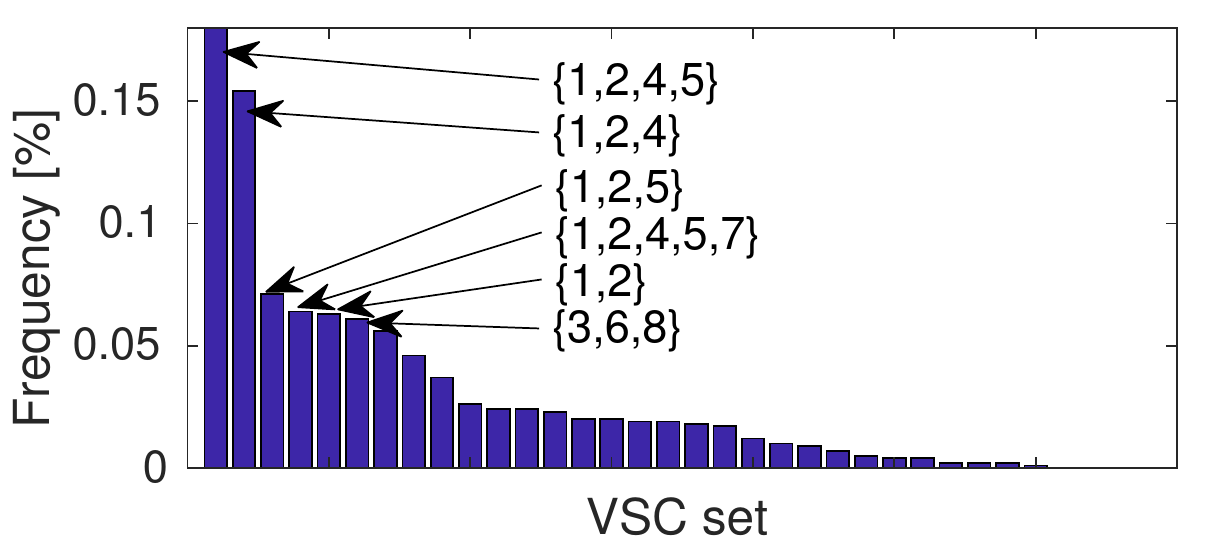}}

\subfloat[]{\includegraphics[width=\columnwidth]{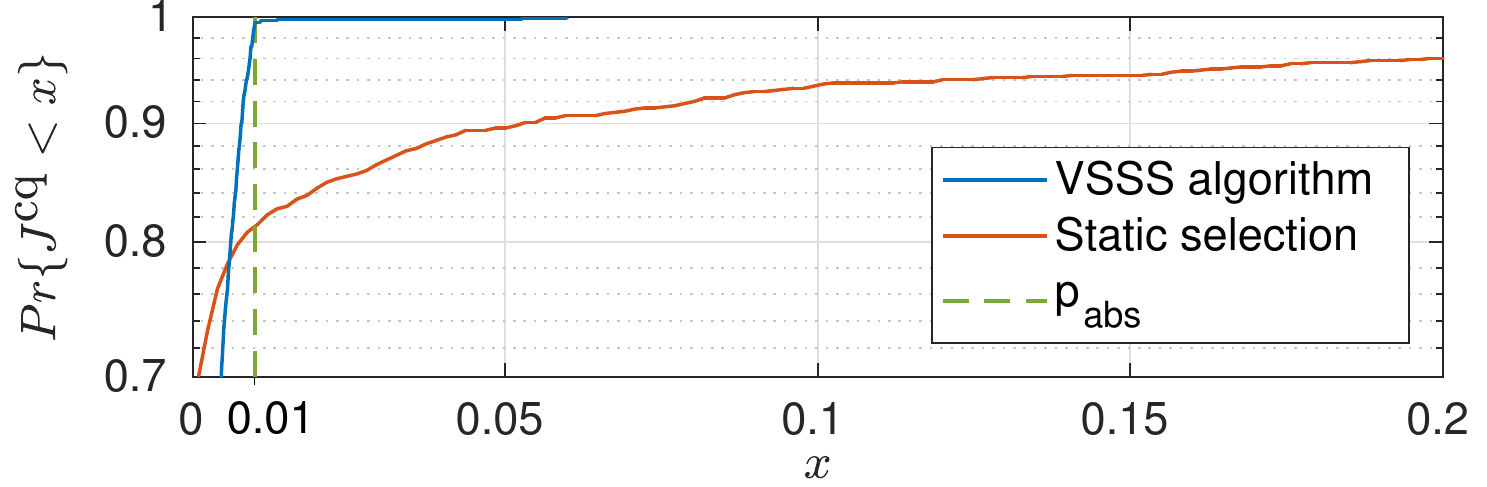}}
    \caption{Application of the proposed selection algorithm to the grid of the case study. (a) Relative frequencies of $\mathcal{V}$ sets selected by Alg.~1. (b) Comparison of the probability of obtaining a $J^\text{cq}$ lower than a target value, with the solution provided by DVSSS and the static selection.}
 \label{fig:realizations}
\end{figure}

In this scenario, the wireless network graph $\mathcal{G}$ is composed of $\omega \geq 1$ static subgraphs, and the vector of generated power, $\mathbf{P}_M$, is stochastic.
We run Alg.~1 over 1000 realizations of $\mathbf{P}_M$.
In Fig.~\ref{fig:realizations}(a) we show a histogram of the sets selected by different iterations of the algorithm, depending on $\mathbf{P}_M$, and in Fig.~\ref{fig:realizations}(b) the probability that $J^\text{cq}$ is lower than a target value $x$, obtained by the DVSSS and by the static selection.
The results show that a dynamic selection provides an increased system reliability with respect to the cost function $J^\text{cq}$, because $J^\text{cq}$ is lower than $p_{\text{abs}}$ in $99.9\%$ of the cases. Secondly, we observe that there are cases in which the static selection outperforms the dynamic selection: the reason is that Alg.~1 selects the set with lowest cardinality that provides the target $J^\text{cq}$, which is not necessarily the one that provides the lowest $J^\text{cq}$.
Finally, it is interesting to observe that, thanks to the presence of the power talk interface, the algorithm does not always select the VSC set in the subgraph $\mathcal{U}_{\omega} = \{ 1,2,4,5,7\}$, but also in $\mathcal{U}_{\omega} = \{ 3,6,8\}$.

\begin{figure}[!tb]
\centering

\subfloat[]{\includegraphics[width=0.95\columnwidth]{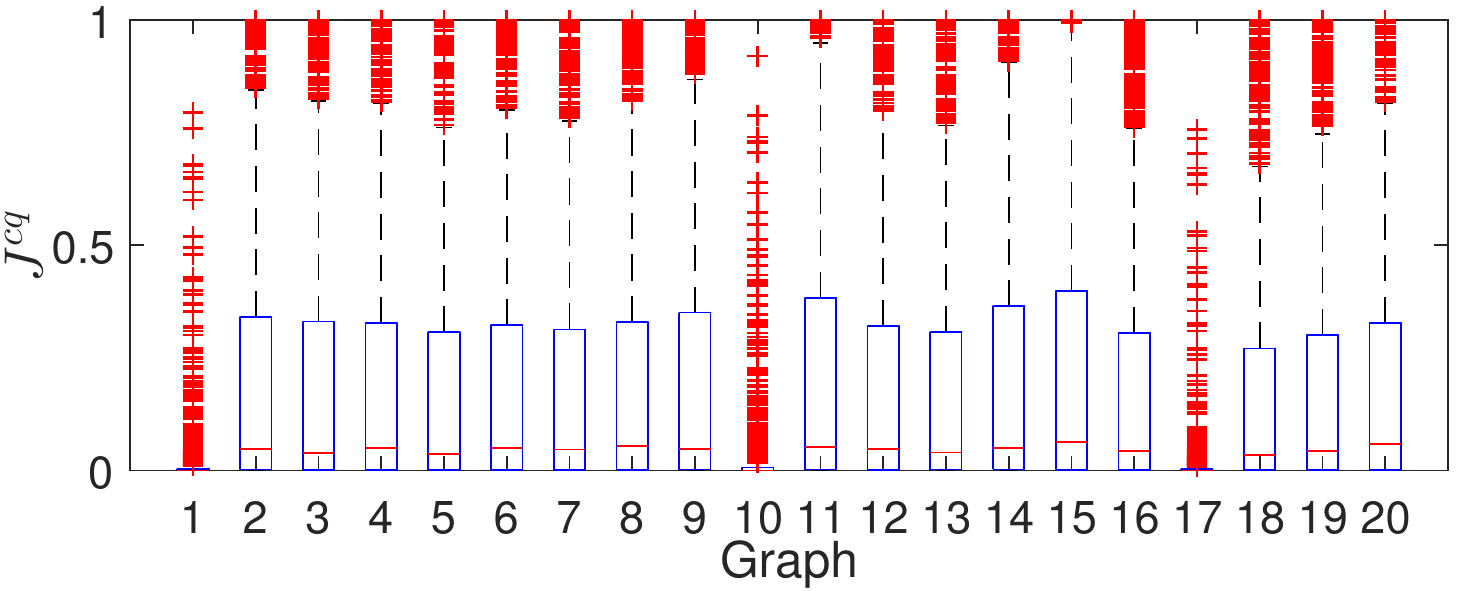}}

\subfloat[]{\includegraphics[width=0.95\columnwidth]{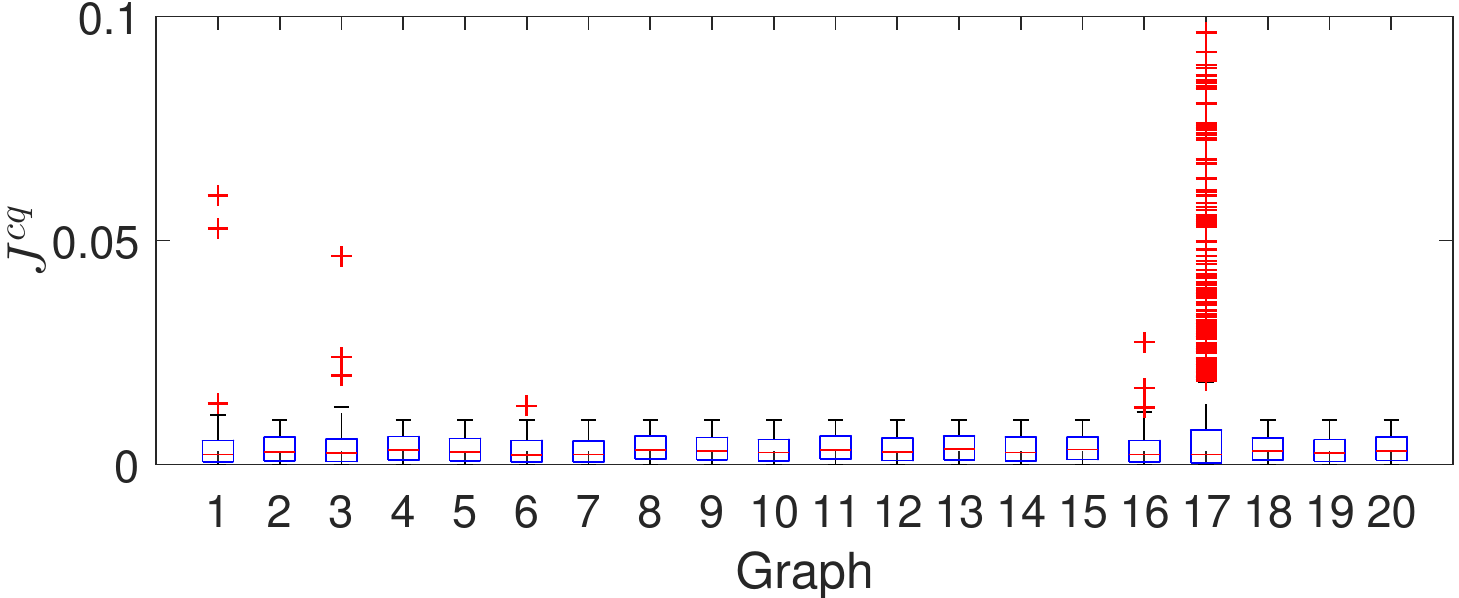}}
    \caption{Boxplots representing the statistics of $J^\text{cq}$ in case of (a) static configuration of $\mathcal{V}$ and (b) when DVSSS is used.}
 \label{fig:boxplots}
\end{figure}

Finally, we verify that the proposed DVSSS improves $J^\text{cq}$ in general, by simulating random wireless network topologies, while keeping the same electrical parameters.
For each $\mathcal{G}$, that is not necessarily connected, we create 1000 realizations of $\mathbf{P}_M$ and power demand $P_d$ and then compare the statistics of DVSSS compared to a static system; for the static selection we used $\mathcal{V}$ that is the most frequently occurring in the DVSSS case.
We generate 20 graphs according to this procedure (graph labeled with 1 is the one of Fig.~\ref{fig:case_study}(b)) and report a boxplot in Fig.~\ref{fig:boxplots} that indicates median value, lower and upper quartiles and outlier values (outside the upper quartile) of $J^\text{cq}$.
The figure shows that the proposed algorithm improves the statistics of $J^\text{cq}$, by finding sets VSC sets that provide $J^\text{cq}< p_\text{abs}$ in most of the cases, except for the case labeled with 17.
The reason is that this graph is split in 4 subgraphs, limiting the possible solutions of Alg.~1, and making the consensus-based control ineffective. The occurrence of this graph warrants further investigation what kind of topologies should be used in order to avoid vast inefficiency.

\section{Conclusion}\label{sec:conclusion}

Motivated by the limitations of the state-of-art distributed secondary control schemes, we have presented a new cyber-pysical architecture tailored for DC MG and exploiting the communication potential residing in the MG electrical equipment.
We have shown that the proposed framework improves the security (robustness against DoS attacks) and control quality (selection of a VSCs set that optimizes a pre-defined metric).
Its cost, quantified as the duration of suspension of the secondary control to permit the information exchange over power talk, is kept low by sending a minimal amount of information over this interface.
Furthermore, this framework promotes the active participation of the MG control in the management of the supporting communication network: the wireless network graph and the control mode are jointly adapted according to the control/communication variations.
As a future work, it is interesting to study the topologies that can benefit most from dynamic reconfiguration and based on that, create suitable recommendations for topology deployment, both for the electrical and the wireless part.

Finally, the proposed architecture is sufficiently general to support additional applications. For example, the information sent over the data plane, i.e., the wireless network, can be encrypted, and the encryption keys be exchanged over the network control plane, to make the data plane traffic decodable only by the PECs, which are the intended users. This implies that MG control layer is not delegating the task of securing the information to the communication interface, but actively contributes to the improvement of the security.

\nocite{*}
\bibliographystyle{IEEEtran}
\bibliography{refs_final}

\begin{thebibliography}{10}
\providecommand{\url}[1]{#1}
\csname url@samestyle\endcsname
\providecommand{\newblock}{\relax}
\providecommand{\bibinfo}[2]{#2}
\providecommand{\BIBentrySTDinterwordspacing}{\spaceskip=0pt\relax}
\providecommand{\BIBentryALTinterwordstretchfactor}{4}
\providecommand{\BIBentryALTinterwordspacing}{\spaceskip=\fontdimen2\font plus
\BIBentryALTinterwordstretchfactor\fontdimen3\font minus
  \fontdimen4\font\relax}
\providecommand{\BIBforeignlanguage}[2]{{%
\expandafter\ifx\csname l@#1\endcsname\relax
\typeout{** WARNING: IEEEtran.bst: No hyphenation pattern has been}%
\typeout{** loaded for the language `#1'. Using the pattern for}%
\typeout{** the default language instead.}%
\else
\language=\csname l@#1\endcsname
\fi
#2}}
\providecommand{\BIBdecl}{\relax}
\BIBdecl

\bibitem{meng2017review}
L.~Meng, Q.~Shafiee, G.~F. Trecate, H.~Karimi, D.~Fulwani, X.~Lu, and J.~M.
  Guerrero, ``Review on control of dc microgrids,'' \emph{IEEE Journal of
  Emerging and Selected Topics in Power Electronics}, 2017.

\bibitem{ref1m}
L.~E. Zubieta, ``Are microgrids the future of energy?: Dc microgrids from
  concept to demonstration to deployment,'' \emph{IEEE Electrification
  Magazine}, vol.~4, no.~2, pp. 37--44, June 2016.

\bibitem{ref2m}
T.~Dragi{\v{c}}evi{\'c}, X.~Lu, J.~C. Vasquez, and J.~M. Guerrero, ``Dc
  microgrids; part i: A review of control strategies and stabilization
  techniques,'' \emph{IEEE Transactions on Power Electronics}, vol.~31, no.~7,
  pp. 4876--4891, July 2016.

\bibitem{ref01}
J.~M. Guerrero, J.~C. Vasquez, J.~Matas, L.~G. De~Vicu{\~n}a, and M.~Castilla,
  ``Hierarchical control of droop-controlled ac and dc microgrids - a general
  approach toward standardization,'' \emph{IEEE Transactions on Industrial
  Electronics}, vol.~58, no.~1, pp. 158--172, 2011.

\bibitem{ref02}
Q.~Shafiee, T.~Dragi{\v{c}}evi{\'c}, P.~Popovski, J.~C. Vasquez, J.~M. Guerrero
  \emph{et~al.}, ``Robust networked control scheme for distributed secondary
  control of islanded microgrids,'' \emph{IEEE Transactions on Industrial
  Electronics}, vol.~61, no.~10, pp. 5363--5374, 2014.

\bibitem{ref03}
P.~Danzi, {\v{C}}.~Stefanovi{\'c}, L.~Meng, J.~M. Guerrero, and P.~Popovski,
  ``On the impact of wireless jamming on the distributed secondary microgrid
  control,'' in \emph{2016 IEEE Globecom Workshops}, Dec 2016, pp. 1--6.

\bibitem{che2014adaptive}
L.~Che, M.~E. Khodayar, and M.~Shahidehpour, ``Adaptive protection system for
  microgrids: Protection practices of a functional microgrid system.''
  \emph{IEEE Electrification magazine}, vol.~2, no.~1, pp. 66--80, 2014.

\bibitem{danzi2016anti}
P.~Danzi, M.~Angjelichinoski, {\v{C}}.~Stefanovi{\'c}, and P.~Popovski,
  ``Anti-jamming strategy for distributed microgrid control based on power talk
  communication,'' in \emph{Communications Workshops (ICC Workshops), 2017 IEEE
  International Conference on}.\hskip 1em plus 0.5em minus 0.4em\relax IEEE,
  2017, pp. 911--917.

\bibitem{ref17}
D.~Kreutz, F.~M. Ramos, P.~E. Verissimo, C.~E. Rothenberg, S.~Azodolmolky, and
  S.~Uhlig, ``Software-defined networking: A comprehensive survey,''
  \emph{Proceedings of the IEEE}, vol. 103, no.~1, pp. 14--76, 2015.

\bibitem{ref3m}
M.~Angjelichinoski, {\v{C}}.~Stefanovi{\'c}, P.~Popovski, H.~Liu, P.~C. Loh,
  and F.~Blaabjerg, ``Power talk: How to modulate data over a dc micro grid bus
  using power electronics,'' in \emph{2015 IEEE Global Communications
  Conference (GLOBECOM)}, Dec 2015, pp. 1--7.

\bibitem{ref5m}
M.~Angjelichinoski, {\v{C}}.~Stefanovi{\'c}, P.~Popovski, H.~Liu, P.~C. Loh,
  and F.~Blaabjer, ``Multiuser communication through power talk in dc
  microgrids,'' \emph{IEEE Journal on Selected Areas in Communications},
  vol.~PP, no.~99, pp. 1--1, 2016.

\bibitem{stefanovic2017resilient}
{\v{C}}.~Stefanovi{\'c}, M.~Angjelichinoski, P.~Danzi, and P.~Popovski,
  ``Resilient and secure low-rate connectivity for smart energy applications
  through power talk in dc microgrids,'' \emph{IEEE Communications Magazine},
  vol.~55, no.~10, pp. 83--89, 2017.

\bibitem{wang2016improved}
P.~Wang, X.~Lu, X.~Yang, W.~Wang, and D.~Xu, ``An improved distributed
  secondary control method for dc microgrids with enhanced dynamic current
  sharing performance,'' \emph{IEEE Transactions on Power Electronics},
  vol.~31, no.~9, pp. 6658--6673, 2016.

\bibitem{cady2015distributed}
S.~T. Cady, A.~D. Dom{\'\i}nguez-Garc{\'\i}a, and C.~N. Hadjicostis, ``A
  distributed generation control architecture for islanded ac microgrids,''
  \emph{IEEE Transactions on Control Systems Technology}, vol.~23, no.~5, pp.
  1717--1735, 2015.

\bibitem{morstyn2016unified}
T.~Morstyn, B.~Hredzak, G.~D. Demetriades, and V.~G. Agelidis, ``Unified
  distributed control for dc microgrid operating modes,'' \emph{IEEE
  Transactions on Power Systems}, vol.~31, no.~1, pp. 802--812, 2016.

\bibitem{liang2013stability}
H.~Liang, B.~J. Choi, W.~Zhuang, and X.~Shen, ``Stability enhancement of
  decentralized inverter control through wireless communications in
  microgrids,'' \emph{IEEE Transactions on Smart Grid}, vol.~4, no.~1, pp.
  321--331, 2013.

\bibitem{ref7m}
H.~Liang, B.~J. Choi, A.~Abdrabou, W.~Zhuang, and X.~S. Shen, ``Decentralized
  economic dispatch in microgrids via heterogeneous wireless networks,''
  \emph{IEEE Journal on Selected Areas in Communications}, vol.~30, no.~6, pp.
  1061--1074, July 2012.

\bibitem{ref12}
T.~Dragi{\v{c}}evi{\'c}, J.~M. Guerrero, J.~C. Vasquez, and D.~{\v{S}}krlec,
  ``Supervisory control of an adaptive-droop regulated dc microgrid with
  battery management capability,'' \emph{IEEE Transactions on Power
  Electronics}, vol.~29, no.~2, pp. 695--706, 2014.

\bibitem{meng2016modeling}
L.~Meng, T.~Dragicevic, J.~Rold{\'a}n-P{\'e}rez, J.~C. Vasquez, and J.~M.
  Guerrero, ``Modeling and sensitivity study of consensus algorithm-based
  distributed hierarchical control for dc microgrids,'' \emph{IEEE Transactions
  on Smart Grid}, vol.~7, no.~3, pp. 1504--1515, 2016.

\bibitem{xia2016probabilistic}
S.~Xia, X.~Luo, K.~W. Chan, M.~Zhou, and G.~Li, ``Probabilistic transient
  stability constrained optimal power flow for power systems with multiple
  correlated uncertain wind generations,'' \emph{IEEE Transactions on
  Sustainable Energy}, vol.~7, no.~3, pp. 1133--1144, 2016.

\bibitem{tarjan1972depth}
R.~Tarjan, ``Depth-first search and linear graph algorithms,'' \emph{SIAM
  journal on computing}, vol.~1, no.~2, pp. 146--160, 1972.

\bibitem{ref4m}
M.~Angjelichinoski, {\v{C}}.~Stefanovi{\'c}, P.~Popovski, and F.~Blaabjerg,
  ``Power talk in dc micro grids: Constellation design and error probability
  performance,'' in \emph{2015 IEEE International Conference on Smart Grid
  Communications (SmartGridComm)}, Nov 2015, pp. 689--694.

\end{thebibliography}

\end{document}